\documentclass[
    reprint,
    nofootinbib,
    amsmath,
    amssymb,
    aps,
    prb,
    longbibliography
]{revtex4-2}

\usepackage{bm}
\usepackage{xcolor}
\usepackage{graphicx}
\usepackage{booktabs}
\usepackage{hyperref}
\usepackage{xurl}

\hypersetup{
    colorlinks=false,
    pdfborder={0 0 1},
    linkbordercolor={1 0 0},
    citebordercolor={1 0 0},
    urlbordercolor={1 0 0}
}

\graphicspath{{draft2_figures_latex/}{figures/}{./}}

\newcommand{\al}{\alpha_{T_{2g}}}
\newcommand{\ket}[1]{\lvert #1\rangle}
\newcommand{\bra}[1]{\langle #1\rvert}
\newcommand{\dd}{\mathrm d}
\newcommand{\sech}{\operatorname{sech}}

\begin{document}

\title{Strain-controlled topology and quantum control at a cubic Ce$^{3+}$ crystal-field crossing}

\author{A. Ghosh}
\affiliation{
Department of Physics and Astronomy,
University of North Florida,
Jacksonville, Florida 32224, USA
}

\author{J. T. Haraldsen}
\affiliation{
Department of Physics and Astronomy,
University of North Florida,
Jacksonville, Florida 32224, USA
}

\date{\today}


\begin{abstract}
Field-tuned crystal-field crossings provide a route to synthetic topology in localized rare-earth degrees of freedom when symmetry-resolved perturbations can independently control the crossing states. Here we show that a field-induced crossing in cubic Ce$^{3+}$ has precisely this structure. For $B\parallel[001]$, the crossing is protected by a twofold rotation, while the $T_{2g}$ shears $\epsilon_{xz}$ and $\epsilon_{yz}$ break that protection and mix the two states through orthogonal pseudospin components. Together with magnetic-field detuning, these controls generate an isolated diabolical point with unit-magnitude Chern charge in a three-dimensional parameter space. Using the CeTe crystal-field scale gives an ideal single-ion crossing near $35.5$ T. A calibrated point-charge calculation gives a representative projected shear coupling $|\alpha_{T_{2g}}|\simeq18.1$ meV per unit tensor strain, corresponding to gaps of $0.036$ and $0.072$ meV for tensor shears of $0.10\%$ and $0.20\%$. We identify the associated elastic and magnetic signatures, discuss the limitations of the electrostatic estimate in a hybridizing Ce compound, and determine the adiabatic and coherence requirements for cyclic geometric control. The result is a symmetry-based framework that connects high-field crystal-field reconstruction, strain response, and parameter-space topology in rare-earth systems.
\end{abstract}

\maketitle

\section{Introduction}

Crystal-electric fields (CEFs) connect the local symmetry of a rare-earth ion to its low-energy magnetic degrees of freedom. For Ce$^{3+}$, the $4f^1$ configuration has a spin--orbit ground multiplet with $J=5/2$. In cubic symmetry, this multiplet splits into a $\Gamma_7$ doublet and a $\Gamma_8$ quartet, and the Stevens-operator formalism provides a compact description of both the zero-field splitting and the perturbations that can shift or mix the resulting states \cite{stevens1952,hutchings1964,Murao1957}. A magnetic field can tune states from different symmetry sectors into degeneracy, while lattice strain provides a separate handle whose effect is fixed by how the corresponding distortion transforms under the local point group.

Two-level degeneracies of parameter-dependent Hamiltonians are naturally described as conical intersections or diabolical points \cite{Herzberg1963,Berry1984,Bruno2006}. When three independent control coordinates couple to the three Pauli matrices, the degeneracy is locally a Weyl point in parameter space and acts as a source of Berry curvature. Closely related topological degeneracies and Berry-phase interference have been studied in molecular magnets \cite{Garg1993,WernsdorferSessoli1999,Bruno2006}, while controlled geometric phases and Chern response have been demonstrated in solid-state qubits and superconducting circuits \cite{Leek2007,Schroer2014,Roushan2014}. The central question for a rare-earth CEF crossing is therefore whether experimentally accessible perturbations span the independent directions needed to control the local two-level Hamiltonian.

Magnetoelastic coupling provides a natural route to such control. To first order in strain, the CEF parameters acquire symmetry-resolved derivatives that determine both the strain dependence of the level scheme and the corresponding elastic response \cite{JensenMackintosh1991,Luthi2005,stoter2021strainCEF}. The broader magnetoelastic literature also makes clear that finite-field ultrasound requires care: rotationally invariant spin--lattice terms can contribute to transverse acoustic modes, as developed by Dohm and Fulde and tested experimentally in TmSb by Wang and L\"uthi \cite{DohmFulde1975,WangLuthi1977,Jensen1988TmSb}. At the same time, static and dynamic strain have become established tools for manipulating quantum degrees of freedom, including rare-earth acoustic dressing and coherent strain control of solid-state spins \cite{Ohta2024Acoustic,Ovartchaiyapong2014,MacQuarrie2015,Barfuss2015}. These developments motivate the question of whether a field-induced CEF crossing can be both determined and controlled through symmetry-selected lattice distortions.

\begin{figure*}[t]
    \centering
    \includegraphics[width=0.94\textwidth]{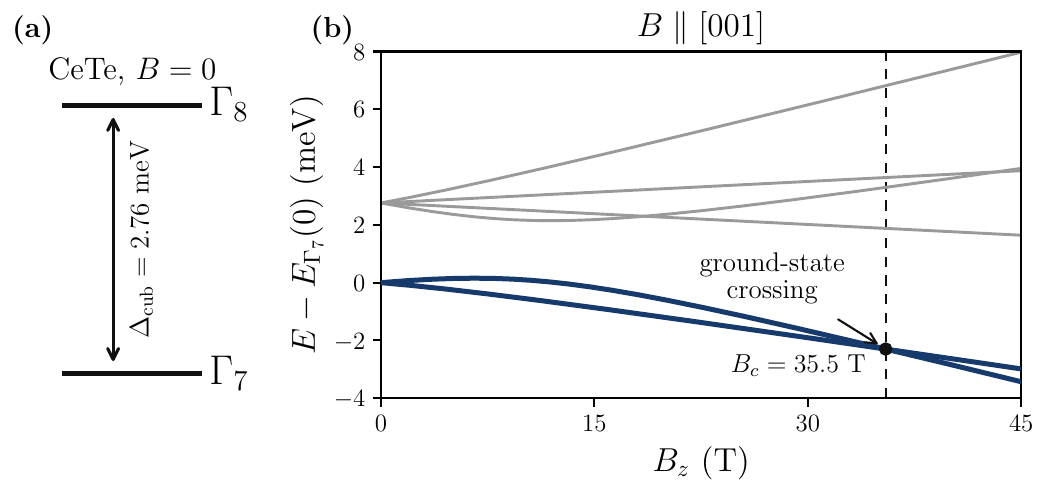}
    \caption{
    Cubic Ce$^{3+}$ spectrum at the CeTe CEF scale.
    (a) The zero-field $\Gamma_7$ doublet and $\Gamma_8$ quartet are separated by $\Delta_{\rm cub}=2.76$ meV.
    (b) Exact six-level spectrum for $B\parallel[001]$, measured from the zero-field $\Gamma_7$ energy. The two lowest branches belong to different $C_{2z}$ sectors and cross at $B_c\simeq35.5$ T.
    }
    \label{fig:spectrum}
\end{figure*}

In this work, we establish this connection for a cubic Ce$^{3+}$ crossing with $B\parallel[001]$. The two crossing states belong to opposite $C_{2z}$ sectors, so the crossing is protected against perturbations that preserve the twofold rotation. The $T_{2g}$ shears $\epsilon_{xz}$ and $\epsilon_{yz}$ are odd under $C_{2z}$ and project onto orthogonal transverse components of the two-level Hamiltonian, while field detuning supplies the longitudinal component. The resulting control space contains an isolated diabolical point with $|C|=1$. We then use CeTe as a quantitative benchmark because its measured $\Gamma_7$--$\Gamma_8$ splitting places the ideal single-ion crossing in the mid-30-T range and its elastic constants already show pronounced CEF signatures \cite{Matsui1988CeTe}. CeTe is nevertheless a correlated magnetic metal with level-dependent exchange and hybridization effects \cite{Nakayama2004CeTe,Matsumura2020Orbital,Kioussis1991}; accordingly, our single-ion construction is intended as a local reference problem rather than a complete high-field phase diagram.

\section{Symmetry-Protected Crossing and Effective Controls}

\subsection{Cubic crossing scale}

Within the isolated $J=5/2$ multiplet, the cubic CEF and Zeeman Hamiltonian is
\begin{equation}
H_0=
B_4\left(O_4^0+5O_4^4\right)
-g_J\mu_B B_zJ_z,
\label{eq:H0}
\end{equation}
with $g_J=6/7$. For $B_4>0$, the zero-field ground state is the $\Gamma_7$ doublet and the $\Gamma_8$ quartet lies higher by $\Delta_{\rm cub}=360B_4$ \cite{stevens1952,hutchings1964,Murao1957}. Sixth-order Stevens operators do not contribute within $J=5/2$ because their rank exceeds $2J$ \cite{hutchings1964}.

For $B\parallel[001]$, two low-energy branches from different symmetry sectors meet at a finite field. Exact diagonalization gives
\begin{equation}
B_c=
\frac{\sqrt{33}}{9}
\frac{\Delta_{\rm cub}}{g_J\mu_B}
\simeq
12.87\,\frac{\mathrm T}{\mathrm{meV}}\Delta_{\rm cub}.
\label{eq:Bc}
\end{equation}
This relation is useful as a simple screening criterion: crossings below $10$, $45$, and $100$ T require $\Delta_{\rm cub}\lesssim0.78$, $3.50$, and $7.77$ meV, respectively. These values are characteristic single-ion field scales rather than predictions for the complete phase diagram of a correlated material.

The crossing is protected by the twofold rotation $C_{2z}=e^{-i\pi J_z}$. At $B_c$, the two states carry opposite eigenvalues, $-i$ and $+i$, so a perturbation that preserves $C_{2z}$ cannot hybridize them. The next CEF level remains separated from the crossing pair by $\Delta_3\simeq1.51\,\Delta_{\rm cub}$, which provides the natural scale controlling the validity of the two-level description. The exact states, block diagonalization, and projection coefficients are given in Appendix~\ref{app:projection}.

For the CeTe value $\Delta_{\rm cub}/k_B\simeq32$ K, Eq.~\eqref{eq:Bc} gives $B_c=35.48$ T and $\Delta_3=4.18$ meV. Figure~\ref{fig:spectrum} shows the full six-level spectrum on this scale.

\subsection{Strain selection rules and the effective Hamiltonian}

The six components of the symmetric strain tensor separate into symmetry channels of the cubic lattice. The off-diagonal shears belong to $T_{2g}$, and their leading quadrupolar coupling may be written in terms of $\{J_i,J_j\}/2$, although a microscopic strain derivative can contain additional CEF operators of the same symmetry \cite{JensenMackintosh1991,Luthi2005,stoter2021strainCEF}.

Under $C_{2z}$, the shears $\epsilon_{xz}$ and $\epsilon_{yz}$ are odd (flipped under transformation), while $\epsilon_{xy}$ and the diagonal strains are even (unchanged under transformation). The two odd shears can therefore hybridize the crossing states at linear order, whereas a sufficiently small $C_{2z}$-preserving deformation can only shift the crossing. Exact projection confirms that $\epsilon_{xz}$ and $\epsilon_{yz}$ act as two orthogonal transverse pseudospin components. Defining the signed projected shear matrix element $\al$ and writing $\delta B=B_z-B_c$, the crossing is described by
\begin{equation}
\begin{aligned}
H_{\rm eff}
&=
\al\left(
\epsilon_{xz}\sigma_x+
\epsilon_{yz}\sigma_y
\right)
-v_B\delta B\,\sigma_z,\\
v_B&=0.0226~\mathrm{meV/T}.
\end{aligned}
\label{eq:Heff}
\end{equation}
The exact projection giving $v_B$ and the relation between $\al$ and the conventional quadrupolar $T_{2g}$ coupling are given in Appendix~\ref{app:projection}.

\begin{figure}
\includegraphics[width=\columnwidth]{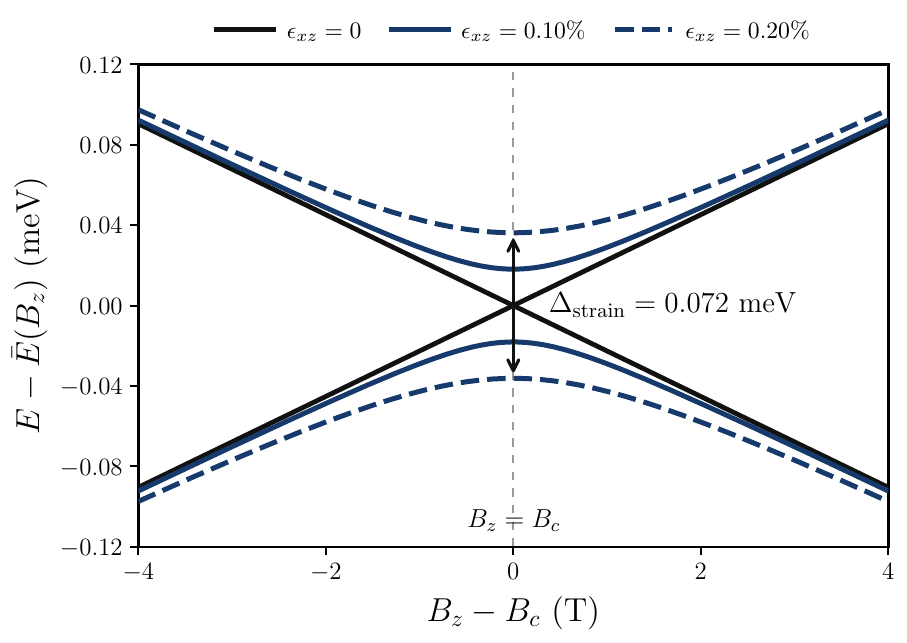}
    \caption{
    Projected avoided crossing for $\epsilon_{yz}=0$ and tensor shears
    $\epsilon_{xz}=0$, $0.10\%$, and $0.20\%$, using the CeTe benchmark
    $|\al|=18.1$ meV per unit tensor strain obtained from the calibrated
    point-charge calculation described in Sec.~\ref{sec:coupling}.
    The corresponding minimum gaps are $0$, $0.0362$, and $0.0724$ meV.
    }
    \label{fig:coupling}
\end{figure}

Equation~\eqref{eq:Heff} contains the central physics of the paper. At the field-tuned crossing, a shear amplitude $\epsilon_\perp=(\epsilon_{xz}^2+\epsilon_{yz}^2)^{1/2}$ opens a gap $2|\al|\epsilon_\perp$. In contrast, symmetry-preserving strain shifts the crossing location without opening a linear-order gap. Figure~\ref{fig:coupling} illustrates the resulting avoided crossing for the CeTe benchmark coupling introduced below.

\subsection{Parameter-space topology and experimental tolerances}

The three coefficients multiplying $\sigma_x$, $\sigma_y$, and $\sigma_z$ in Eq.~\eqref{eq:Heff} are independently controlled by $\epsilon_{xz}$, $\epsilon_{yz}$, and $\delta B$. Because the linear map from these physical coordinates to the pseudospin field is nonsingular when $\al\neq0$, the degeneracy at the origin is isolated. It is therefore a diabolical point, or equivalently a Weyl point in the synthetic three-dimensional control space, with unit-magnitude Chern charge. The sign assigned to a particular band depends on orientation and band conventions, while the invariant statement is $|C|=1$ \cite{Berry1984,Bruno2006}. Appendix~\ref{app:projection} gives the explicit Jacobian and Chern-sign convention.

A transverse magnetic field breaks the same protecting symmetry and therefore acts as an additional transverse control. The projected transverse-field gap is $0.11\,B_\perp$ meV when $B_\perp$ is measured in tesla. At the CeTe-scale crossing, a $0.1^\circ$ field misalignment therefore produces a gap of approximately $6.85~\mu$eV; keeping this contribution below $10~\mu$eV requires alignment to about $0.15^\circ$. Residual shear and transverse field are components of the same effective pseudofield and should be treated vectorially rather than as independent scalar broadenings.

The local two-level model is controlled by the separation $\Delta_3$ to the remaining CEF levels. For the curves shown here, the strain and detuning energies remain at least an order of magnitude below $\Delta_3$, and direct six-level checks agree closely with the projected spectrum. Numerical details are given in Appendix~\ref{app:projection}. This separation is important because it allows the symmetry and topology of the crossing to be discussed independently of the material-specific corrections that determine its precise experimental realization.

\section{Material and Coupling Scales}

\subsection{CeTe and candidate field scales}

\begin{figure}[t]
    \centering
    \includegraphics[width=\columnwidth]{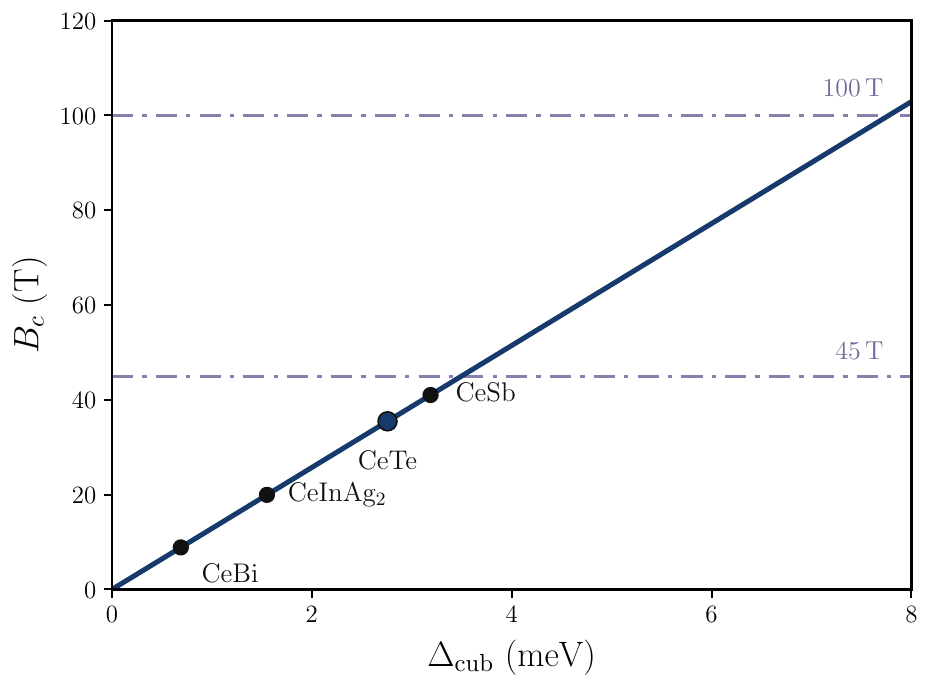}
    \caption{
    Ideal cubic single-ion crossing field as a function of the zero-field CEF splitting, with
    $B_c\simeq12.87\,\Delta_{\rm cub}$ T when $\Delta_{\rm cub}$ is expressed in meV.
    The markers use representative reported CEF splittings for CeBi, CeInAg$_2$, CeTe, and CeSb
    \cite{Pierre1981CeInAg2,RouraBas2007CEF,Matsui1988CeTe}.
    Horizontal lines indicate representative $45$ T dc and $100$ T pulsed-field capabilities demonstrated experimentally
    \cite{Hahn2019HighField,Jaime2012HighField}.
    The plotted values are screening estimates, not complete high-field phase-diagram predictions.
    }
    \label{fig:materials}
\end{figure}

Equation~\eqref{eq:Bc} turns the measured zero-field CEF splitting into a useful first-pass field scale. Representative reported splittings for CeBi, CeInAg$_2$, CeTe, and CeSb are approximately $8$, $18$, $32$, and $37$ K, respectively \cite{Pierre1981CeInAg2,RouraBas2007CEF,Matsui1988CeTe}. In the ideal cubic single-ion model, these correspond to characteristic crossing fields of approximately $8.9$, $20.0$, $35.5$, and $41.0$ T, as summarized in Fig.~\ref{fig:materials}. Exchange, hybridization, magnetic order, and departures from the ideal local symmetry can all shift or reconstruct these scales in a real compound.

CeTe is a useful proof-of-principle benchmark because its $C_{44}$ and $(C_{11}-C_{12})/2$ modes show clear CEF-related softening associated with the $\Gamma_7$--$\Gamma_8$ level scheme \cite{Matsui1988CeTe}. High-field ultrasound has already followed its low-temperature phase boundaries to $28$ T \cite{Nakamura2020CeTe}. At the same time, CeTe orders antiferromagnetically near $2.2$ K and exhibits orbital-dependent exchange, so the experimentally observed evolution near $35$ T need not coincide with the ideal isolated-ion spectrum \cite{Nakayama2004CeTe,Matsumura2020Orbital}. This is precisely why the symmetry distinction between gapping and non-gapping strain channels is useful; it can be tested even when collective interactions renormalize the field scale.

\subsection{Projected shear-coupling estimate}
\label{sec:coupling}

To obtain a concrete strain scale, we use a point-charge model of the rock-salt CeTe environment implemented with \texttt{PyCrystalField} \cite{scheie2021pycrystalfield}. This approach follows the standard magnetoelastic crystal-field construction in which strain-dependent coupling constants are obtained from derivatives of the CEF parameters with respect to symmetry-resolved lattice distortions \cite{delMoralLee1980PointCharge,stoter2021strainCEF}. Point-charge calculations of this type have previously been used to estimate magnetoelastic coefficients in rare-earth compounds, with quantitative agreement for some cubic systems \cite{Stanley1985ErAl2}. More recently, strained crystal structures have also been treated directly within \texttt{PyCrystalField} point-charge calculations \cite{Akatsuka2024DyTe3}. Here, the zero-strain electrostatic scale is first calibrated to the measured $\Delta_{\rm cub}/k_B = 32$ K and then held fixed while the lattice is sheared. The detailed finite-difference procedure, shell convergence, and symmetry checks are given in Appendix~\ref{app:pc}.

Across four tested effective charge ratios, the large-cutoff values of the projected matrix element range from $17.6$ to $18.6$ meV per unit tensor strain. We therefore use $\al$ = 18.1 meV per unit tensor strain as a representative benchmark. In the leading single-quadrupole parameterization, this corresponds to $|g_{T_{2g}}^{\rm eff}|\simeq8.48$ meV per unit tensor strain. The spread across the electrostatic models is small, but it should not be interpreted as a complete material uncertainty.

\begin{table}[t]
    \caption{
    Calibrated point-charge estimates of the projected shear
    coupling. For each effective charge ratio,
    $|\al^{\rm PC}|$ is the mean over the large-cutoff tail,
    while the tail standard deviation quantifies the residual
    variation between successive complete-shell cutoffs. All energy entries are in meV per
    unit tensor strain.
    }
    \label{tab:pc}
    \centering
    \begin{tabular}{@{}cccc@{}}
        \toprule
        $q_{\rm Ce}/|q_{\rm Te}|$
        & $|\al^{\rm PC}|$
        & Spread
        & $|g^{\rm eff}_{T_{2g}}|$ \\
        \midrule
        0.0 & 17.61 & 0.11 & 8.25 \\
        0.5 & 17.95 & 0.11 & 8.41 \\
        1.0 & 18.27 & 0.12 & 8.56 \\
        1.5 & 18.56 & 0.13 & 8.69 \\
        \bottomrule
    \end{tabular}
\end{table}

This limitation is especially important for Ce compounds. Work on Ce monopnictides shows that band--$f$ and $p$--$f$ hybridization can strongly renormalize the apparent CEF splitting relative to a purely electrostatic picture \cite{TakahashiKasuya1985,WillsCooper1987}. Calculations spanning CeSb to CeTe likewise find a strong dependence of the hybridization-dressed CEF scale and exchange on the chemical environment \cite{Kioussis1991}. More generally, calculations for rare-earth intermetallics show that conduction-electron screening can substantially renormalize point-charge magnetoelastic coefficients and, for some symmetry channels, can even alter their sign \cite{delMoral1983Screening}. Calibrating a point-charge model to the measured zero-strain splitting therefore does not guarantee that its strain derivative is quantitatively exact. We use $18.1$ meV per unit tensor strain as a physically motivated reference scale, while the material-specific value of $|\al|$ should ultimately be determined by strain-dependent spectroscopy, ultrasound, or a microscopic electronic-structure calculation.

At this benchmark, tensor shears of $0.10\%$ and $0.20\%$ open gaps of $36.2$ and $72.4~\mu$eV. Cryogenic piezoelectric platforms have demonstrated continuously tunable sub-percent strain \cite{Hicks2014Strain}, and rare-earth quadrupolar systems such as TmVO$_4$ exhibit strong strain susceptibility, which can be probed via elastocaloric measurements \cite{Zic2024}. These precedents make the static strain scale realistic enough to motivate a direct experimental test, while the dynamical requirements for coherent cyclic control are more restrictive and are treated separately below.

\section{Experimental Signatures of the Crossing}
\label{sec:signatures}

The effective Hamiltonian predicts several observables that can be tested before any coherent-control experiment is attempted. The most direct is spectroscopy under controlled shear; an odd $T_{2g}$ shear should open an avoided crossing with a gap linear in $|\epsilon_\perp|$, whereas a $C_{2z}$-preserving strain should primarily shift the crossing field. This symmetry contrast is more robust than the absolute point-charge value of $\al$ and therefore provides a clean experimental test of the projected Hamiltonian.

The same matrix element that opens the strain gap also produces a $T_{2g}$ elastic susceptibility. In the equilibrium two-level limit, the electronic correction to $c_{44}$ is negative and becomes sharply enhanced near the crossing on cooling. For the CeTe density and the benchmark $|\al|=18.1$ meV, the ideal crossing-scale result is
\begin{equation}
\delta c_{44}(B_c,T)\simeq-\frac{2.37~\mathrm{GPa\,K}}{T}.
\end{equation}
The full thermodynamic derivation is given in Appendix~\ref{app:response}. Such CEF-induced elastic softening and magnetoelastic anomalies have a long history in rare-earth systems \cite{Luthi2005,Morin1995}. The divergence of the isolated two-level expression should not be extrapolated through a loss of lattice stability. Sufficiently strong coupling can instead favor a cooperative distortion that self-gaps or otherwise reconstructs the ideal crossing.

For quantitative finite-frequency ultrasound in a magnetic field, the equilibrium strain susceptibility is not enough to explain the whole story. Rotational invariance introduces additional spin--lattice terms, and classic TmSb measurements showed that the same nominal $c_{44}$ mode can split when the propagation and polarization geometry relative to the field is interchanged \cite{DohmFulde1975,WangLuthi1977}. Later analysis showed that dipolar and finite-deformation effects can also be important \cite{Jensen1988TmSb}. A high-field ultrasound test of the present proposal should use static two-level softening as the symmetry-resolved baseline, while treating rotational and dynamical corrections within a geometry-consistent magnetoelastic theory.

\begin{figure*}[t]
    \centering
    \includegraphics[width=0.94\textwidth]{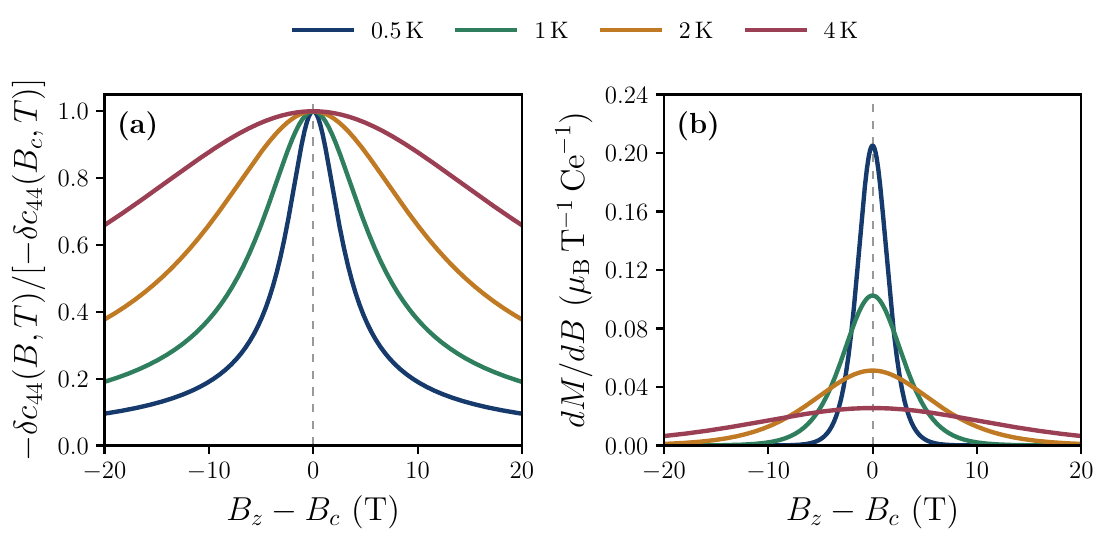}
    \caption{
    Ideal two-level responses for $T=0.5$, $1$, $2$, and $4$ K.
    (a) Normalized magnitude of the $T_{2g}$ elastic softening around the crossing.
    (b) Crossing contribution to the differential magnetization. Its integral through the reconstructed crossing is $0.78137\,\mu_B$ per Ce in the linear two-level model.
    The plotted curves show the equilibrium local response; quantitative ultrasound in field can contain additional rotational and dynamical magnetoelastic contributions.
    }
    \label{fig:response}
\end{figure*}

The crossing also reconstructs the longitudinal magnetic moment. The two exact crossing states differ by
$\Delta M$ = 0.78 $\mu_B$ per Ce, so the ideal zero-shear model gives a step-like change that is thermally broadened around $B_c$. Figure~\ref{fig:response} summarizes both this magnetic response and the corresponding $T_{2g}$ elastic softening. Field-induced CEF level crossings with strong magnetic signatures are not unique to the present construction. HoVO$_4$ exhibits a high-field level crossing accompanied by a magnetization jump and magnetoelastic response \cite{Morin1995}, while recent field-dependent optical spectroscopy of CsErSe$_2$ directly resolves a ground-state CEF crossing associated with a sharp magnetic transition \cite{Whitelock2025}. These systems provide useful experimental context for separating single-ion reconstruction from the collective behavior that can emerge around it.

Taken together, spectroscopy, ultrasound, and magnetization provide a staged way to test the proposal. The first goal is to locate the actual high-field reconstruction and determine whether odd and even strain channels behave as predicted. The second is to extract the real material value of $|\al|$ and determine whether the crossing is preempted or strongly renormalized by exchange or lattice instability. Elastocaloric response provides an additional possible thermodynamic probe of strong quadrupole--strain susceptibility \cite{Zic2024}. Only after this static structure is established does it become useful to ask whether the same two-level system can be manipulated coherently.

\section{Cyclic Shear and Geometric Control}

Driving the two odd shears in quadrature,
$\epsilon_{xz}(t)=\epsilon_0\cos\phi(t)$ and
$\epsilon_{yz}(t)=\epsilon_0\sin\phi(t)$,
causes the effective pseudofield to encircle the degeneracy. With
$R=|\al|\epsilon_0$ and $m=-v_B\delta B$, the instantaneous gap remains
$2\sqrt{R^2+m^2}$. For a loop with circulation $s=\pm1$, the adiabatic geometric phase of the two eigenstates is
\begin{equation}
\gamma_\pm(s)
=
\mp s\pi
\left(
1-\frac{m}{\sqrt{R^2+m^2}}
\right)
\pmod{2\pi}.
\label{eq:Berry}
\end{equation}
The derivation is given in Appendix~\ref{app:berry}.

At exact centering, each branch acquires a Berry phase of magnitude $\pi$. The relative geometric phase between the two branches is then $2\pi$, so a perfectly centered loop is topologically nontrivial but does not by itself generate a nontrivial relative phase within the two-level subspace. Finite detuning continuously tunes the relative geometric phase. This distinction is useful experimentally; the Chern charge is diagnosed by the flux structure of the degeneracy, while a geometric phase gate or interferometric signal depends on the actual loop geometry. Berry-phase interference in magnetic systems and controlled geometric phases in solid-state qubits provide direct conceptual precedents \cite{WernsdorferSessoli1999,Bruno2006,Leek2007}.

\begin{figure*}[t]
    \centering
    \includegraphics[width=0.96\textwidth]{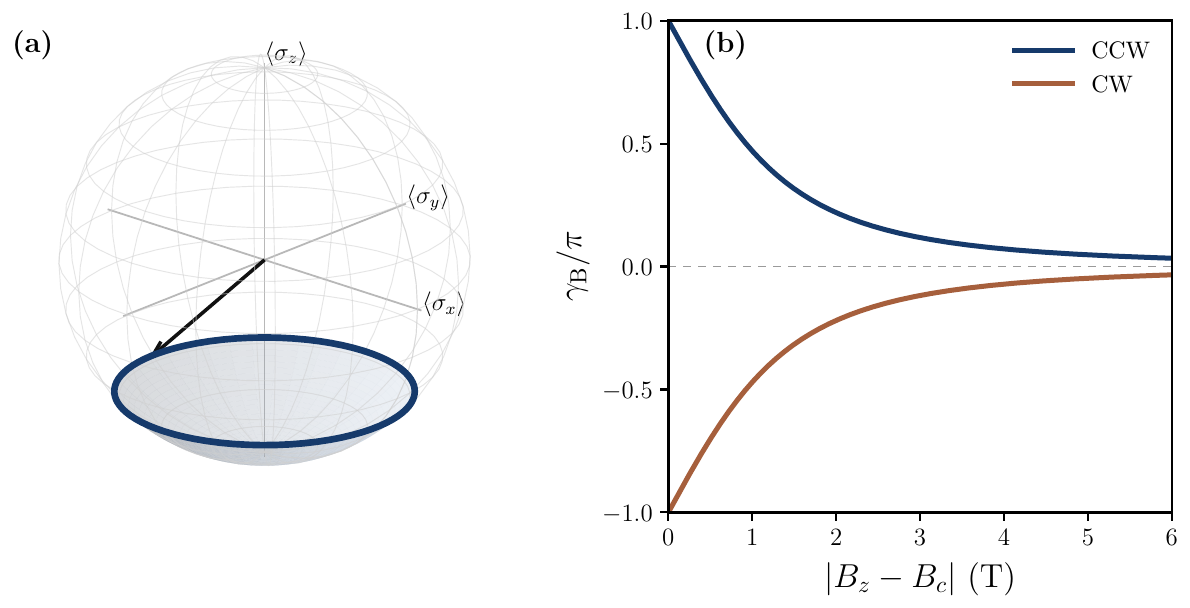}
    \caption{
    Geometric phase from circular shear control.
    (a) Bloch-sphere trajectory of the lower eigenstate for a circular $T_{2g}$ shear drive. The displayed example uses $|\al|=18.1$ meV per unit tensor strain, $\epsilon_0=0.20\%$, and $|B_z-B_c|=1.5$ T.
    (b) Lower-state Berry phase for counterclockwise (CCW) and clockwise (CW) loops on the $B_z\leq B_c$ side. Circulation is defined as viewed from positive $d_z$, and phases are defined modulo $2\pi$.
    }
    \label{fig:geometry}
\end{figure*}

An interferometric protocol must also remove or calibrate the dynamical phase. Equal-duration clockwise and counterclockwise loops with identical energy histories acquire the same dynamical phase and opposite geometric phases, making their comparison a natural phase-sensitive measurement \cite{Jones2000Geometric}. Figure~\ref{fig:geometry} shows the corresponding Bloch-sphere trajectory and the Berry phase versus detuning.

Adiabatic control requires the loop to be slow relative to the instantaneous gap, while coherent readout requires it to remain short compared with the dephasing time $T_2$. For a centered circular loop, requiring an adiabaticity parameter $\eta_{\rm ad}\leq0.01$ gives the minimum durations listed in Table~\ref{tab:drive}. The important point is that a high acoustic carrier frequency does not automatically imply adiabatic encircling. Strain amplitude, loop frequency, and detuning enter together. The detailed nonadiabatic coupling is derived in Appendix~\ref{app:berry}.

\begin{table}[t]
    \caption{
    Centered-loop adiabaticity benchmarks for
    $|\al|=18.1$ meV per unit tensor strain. The final column gives the
    minimum loop duration required to satisfy
    $\eta_{\rm ad}\leq0.01$.
    }
    \label{tab:drive}
    \centering
    \begin{tabular}{@{}cccc@{}}
        \toprule
        $\epsilon_0$
        & $2R$ ($\mu$eV)
        & $2R/h$ (GHz)
        & $T_{\rm loop}^{\min}$ (ns) \\
        \midrule
        $10^{-5}$        & 0.362 & 0.0875 & 571 \\
        $10^{-4}$        & 3.62  & 0.875  & 57.1 \\
        $2\times10^{-3}$ & 72.4  & 17.5   & 2.86 \\
        \bottomrule
    \end{tabular}
\end{table}

Mechanical quantum control at comparable conceptual levels has already been demonstrated in other solid-state platforms. Strain coupling of individual NV spins to mechanical resonators has been measured directly \cite{Ovartchaiyapong2014}, and mechanically driven Rabi and Ramsey protocols have been realized using resonant stress \cite{MacQuarrie2015}. Strong coherent mechanical driving of a single spin has also been demonstrated \cite{Barfuss2015}. Rare-earth acoustic dressing provides a complementary precedent closer to the present microscopic setting \cite{Ohta2024Acoustic}. These results establish that strain can, in principle, act as a coherent quantum-control field. A viable experiment therefore requires both the adiabatic lower bound on $T_{\rm loop}$ and the independent condition $T_{\rm loop}\ll T_2$ under the actual field, temperature, and drive conditions.

\section{Discussion and Conclusion}

The central result is a symmetry statement that holds regardless of the details of the material realization. A cubic Ce$^{3+}$ crossing for $B\parallel[001]$ is protected by $C_{2z}$, and the two odd $T_{2g}$ shears provide the two independent transverse controls needed to gap it. Together with magnetic-field detuning, they span the full local pseudospin space and generate an isolated degeneracy in parameter space with $|C|=1$. Transverse magnetic fields realize the same topology, but strain is especially useful because its symmetry can be selected experimentally and because the same coupling is visible through the elastic response.

CeTe provides a concrete scale. Its measured CEF splitting places the ideal crossing near $35.5$ T, and the calibrated point-charge model suggests that sub-percent shear could open gaps on the order of tens of $\mu$eV. At the same time, the known sensitivity of Ce CEF scales to hybridization and exchange means that the point-charge value of $|\al|$ should be regarded as a benchmark \cite{TakahashiKasuya1985,WillsCooper1987,Kioussis1991}. The experimentally decisive quantities are the actual field of the reconstruction, the contrast between odd and symmetry-preserving strain channels, and the measured slope of the strain-induced gap.

This naturally suggests an experimental hierarchy. Static high-field spectroscopy or magnetization can first locate the crossing. Controlled $T_{2g}$ strain can then test whether the crossing gaps vary linearly, whereas even strains primarily shift them. Ultrasound can probe the associated elastic susceptibility, provided rotationally invariant magnetoelastic contributions are included in the finite-frequency analysis. If the crossing pair remains coherent on a sufficiently long timescale, two-quadrature shear can then be used to encircle the degeneracy and measure its geometric phase.

More broadly, the same logic can be applied to other rare-earth ions and crystal symmetries. The ingredients are a field-tunable low-energy degeneracy, a protecting symmetry, and experimentally accessible perturbations whose irreducible representations project onto independent pseudospin directions. Framed this way, the problem becomes a materials-design question. One can identify the crossing, classify the allowed strain channels, and determine whether those controls span the local two-level Hamiltonian. This provides a route from conventional crystal-field spectroscopy to mechanically addressable synthetic topology in localized quantum systems.

\section*{Acknowledgments}

AG and JTH acknowledge support from the Institute for Materials Science at Los Alamos National Laboratory. AI tools were used to assist with grammatical editing of the manuscript. However, all scientific content, calculations, interpretations, and conclusions were conducted, reviewed, and verified by the authors.

\appendix

\section{Exact crossing states, projection, and topology}
\label{app:projection}

In the ordered subspaces $\{\ket{5/2},\ket{-3/2}\}$ and
$\{\ket{3/2},\ket{-5/2}\}$, the two relevant blocks of Eq.~\eqref{eq:H0} are
\begin{align}
H_1
&=
\begin{pmatrix}
60B_4-\frac52E_b & 60\sqrt5B_4 \\
60\sqrt5B_4 & -180B_4+\frac32E_b
\end{pmatrix},
\label{eq:blocks1}
\\
H_2
&=
\begin{pmatrix}
-180B_4-\frac32E_b & 60\sqrt5B_4 \\
60\sqrt5B_4 & 60B_4+\frac52E_b
\end{pmatrix},
\label{eq:blocks2}
\end{align}
where $E_b=g_J\mu_BB_z$. Equating the lower eigenvalues gives the zero-field Kramers root and the positive finite-field root
\[
E_b^c=40\sqrt{33}B_4.
\]
At the finite-field crossing, the common energy is $E_c=-540B_4$. The next level lies above the pair by
\[
\Delta_3
=
\frac{33-\sqrt{33}}{18}\Delta_{\rm cub}
\simeq1.51419\,\Delta_{\rm cub}.
\]

Set $u=\sqrt{33}+6$. A real normalized basis that fixes the phase convention is
\begin{equation}
\begin{aligned}
\ket{\psi_1}
&=
\frac{-\sqrt5\,u\ket{5/2}/5+\ket{-3/2}}
{\sqrt{1+u^2/5}},
\\
\ket{\psi_2}
&=
\frac{-\sqrt5\,u\ket{3/2}/3+\ket{-5/2}}
{\sqrt{1+5u^2/9}}.
\end{aligned}
\label{eq:states}
\end{equation}
They satisfy
$C_{2z}\ket{\psi_1}=-i\ket{\psi_1}$ and
$C_{2z}\ket{\psi_2}=+i\ket{\psi_2}$.

For the leading quadrupolar $T_{2g}$ operators
\[
O_{ij}=\frac12\{J_i,J_j\},
\]
direct projection gives
\begin{equation}
\begin{aligned}
PO_{xz}P&=a\sigma_x,
&
PO_{yz}P&=a\sigma_y,
\\
PO_{xy}P&=0,
&
PJ_zP&=j_0I+c_z\sigma_z,
\end{aligned}
\label{eq:projection}
\end{equation}
with
\[
a=\frac{5\sqrt{5973}}{181},
\qquad
j_0=\frac{60\sqrt{33}}{181},
\qquad
c_z=\frac{165}{362}.
\]
We define
\[
\al=\bra{\psi_1}V_{xz}\ket{\psi_2},
\]
where $V_{xz}$ is the derivative of the CEF Hamiltonian with respect to the tensor shear $\epsilon_{xz}=\epsilon_{zx}$. Cubic symmetry fixes the $yz$ derivative to the orthogonal imaginary pseudospin component. In the single-quadrupole approximation, $\al=ag_{T_{2g}}$.

The Zeeman perturbation projects as
\[
-g_J\mu_B\delta B\,PJ_zP
=
-g_J\mu_Bj_0\delta B\,I
-v_B\delta B\,\sigma_z,
\]
where
\[
v_B=c_zg_J\mu_B=0.0226~\mathrm{meV/T}.
\]
After dropping the common identity term, this reproduces Eq.~\eqref{eq:Heff}.

A transverse magnetic field also breaks $C_{2z}$. The exact matrix elements are
\[
PJ_xP=
\frac{15}{\sqrt{181}}\sigma_x,
\qquad
PJ_yP=
\frac{15}{\sqrt{181}}\sigma_y
\]
within the same phase convention. Defining
$\beta=g_J\mu_B(15/\sqrt{181})$, the projected transverse Zeeman term is
$-\beta(B_x\sigma_x+B_y\sigma_y)$, giving
\[
\Delta_{B_\perp}=2\beta B_\perp
=0.11064\,B_\perp~\mathrm{meV/T}.
\]

For the physical coordinates $(\epsilon_{xz},\epsilon_{yz},\delta B)$,
\[
\mathbf d=
(\al\epsilon_{xz},\al\epsilon_{yz},-v_B\delta B),
\qquad
H_{\rm eff}=\mathbf d\cdot\boldsymbol\sigma.
\]
The linear map has
\begin{equation}
\det\left[
\frac{\partial\mathbf d}
{\partial(\epsilon_{xz},\epsilon_{yz},\delta B)}
\right]
=-\al^2v_B\neq0,
\label{eq:jacobian}
\end{equation}
so the degeneracy is isolated. With the Berry connection
$\mathcal A_\pm=i\bra{u_\pm}\nabla_\lambda u_\pm\rangle$ and the outward orientation in the physical control coordinates, the convention above gives $C_-=-1$ and $C_+=+1$. Reversing the orientation or exchanging bands reverses the sign, while $|C|=1$ is invariant \cite{Berry1984,Bruno2006}.

The projected description is valid when the perturbation and thermal scales remain small compared with $\Delta_3$. A useful condition is
\[
|\al|\epsilon_\perp,\qquad
v_B|\delta B|,\qquad
k_BT
\ll\Delta_3,
\]
together with the more conservative requirement
$\|\delta H\|\ll\Delta_3$ when couplings to the other CEF states are included. For the CeTe-scale curves in the main text,
$|\al|\epsilon_\perp\leq0.0362$ meV and
$v_B|\delta B|\leq0.0905$ meV over $|\delta B|\leq4$ T, compared with
$\Delta_3=4.18$ meV. Direct six-level diagonalization over this window differs from the projected pair half-splitting by less than
$1.8\times10^{-3}$ meV.

\section{Point-charge strain derivative}
\label{app:pc}

The point-charge estimate is constructed from the rock-salt CeTe environment using
$a_{\rm lattice}=6.361$ \AA. At each complete shell cutoff, a single electrostatic scale is chosen to reproduce the measured zero-strain cubic splitting and is then held fixed under deformation. This distinction is essential: recalibrating the charge separately for each strained geometry would remove part of the strain response from the estimate.

For a homogeneous tensor shear,
\[
\mathbf r'=(I+\boldsymbol\epsilon)\mathbf r,
\qquad
\epsilon_{xz}=\epsilon_{zx}=\eta.
\]
The strain derivative is evaluated by the central difference
\begin{equation}
V_{xz}
\simeq
\frac{H(+\eta)-H(-\eta)}{2\eta},
\label{eq:pc_derivative}
\end{equation}
and projected onto the exact unstrained crossing states,
\[
|\al^{\rm PC}|
=
\left|
\bra{\psi_1}V_{xz}\ket{\psi_2}
\right|,
\]
where PC denotes the point-charge estimate.
The calculation reproduces the expected symmetry structure: the $xz$ derivative is a real transverse matrix element, the $yz$ derivative gives the orthogonal imaginary component with the same magnitude, and the $xy$ derivative vanishes in the crossing subspace.

\begin{figure}
    \centering
    \includegraphics[width=\columnwidth]{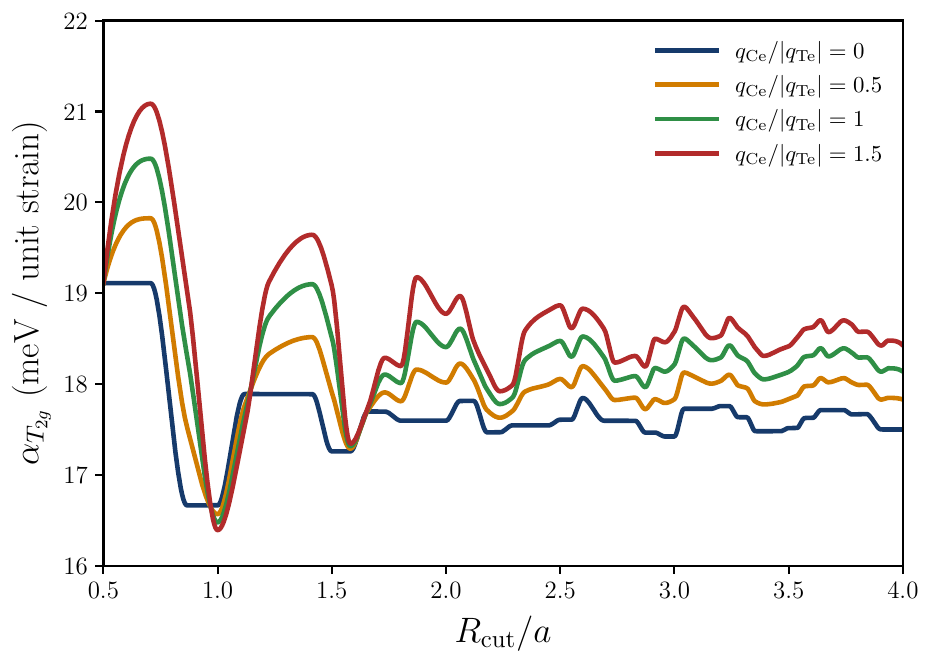}
    \caption{
    Point-charge estimate of $|\al|$ versus the complete-shell cutoff $R_{\rm cut}/a_{\rm lattice}$ for the indicated effective charge ratios. Each cutoff is calibrated to the same zero-strain cubic splitting before evaluating the strain derivative at fixed charge scale. Smooth curves are guides to the eye through the discrete complete-shell calculations. The large-cutoff values settle in the $17.6$--$18.6$ meV per unit tensor-strain range, with $18.1$ meV used as the main-text benchmark.
    }
    \label{fig:pc}
\end{figure}

Only complete radial coordination shells are used. The residual shell oscillations motivate reporting a large-cutoff average and shell-to-shell spread separately. These quantities measure numerical sensitivity within the chosen electrostatic model and do not include uncertainty from screening, hybridization, internal relaxation, or strain-dependent charge redistribution. In particular, the known hybridization sensitivity of Ce CEF splittings \cite{TakahashiKasuya1985,WillsCooper1987,Kioussis1991} is not represented by the narrow spread in Fig.~\ref{fig:pc}.

\section{Thermodynamic response functions}
\label{app:response}

\subsection{Elastic response}

For one tensor shear $\epsilon=\epsilon_{xz}$, the projected Hamiltonian can be written
\[
H=m\sigma_z+\al\epsilon\sigma_x,
\qquad
m=-v_B\delta B.
\]
The two eigenvalues are $\pm E$, with
\[
E=\sqrt{m^2+\al^2\epsilon^2}.
\]
The two-level free energy per Ce ion is
\[
F(\epsilon)
=
-k_BT
\ln\left[
2\cosh\left(\frac{E}{k_BT}\right)
\right].
\]
Differentiating twice with respect to tensor strain gives
\[
\left.
\frac{\partial^2F}{\partial\epsilon^2}
\right|_{\epsilon=0}
=
-\al^2
\frac{\tanh\left(|m|/k_BT\right)}{|m|}.
\]

The conventional Voigt modulus $c_{44}$ is defined using engineering shear
$\gamma=2\epsilon$, for which
\[
f_{\rm el}
=
\frac12c_{44}\gamma^2
=
2c_{44}\epsilon^2.
\]
The electronic correction is therefore
\begin{equation}
\delta c_{44}
=
-\frac{n_{\rm Ce}\al^2}{4}
\frac{\tanh\left(|m|/k_BT\right)}{|m|},
\label{eq:c44_app}
\end{equation}
where the factor of $1/4$ converts the tensor-strain derivative to the conventional shear modulus. At the crossing,
\[
\delta c_{44}(B_c,T)
=
-\frac{n_{\rm Ce}\al^2}{4k_BT}.
\]
Using $n_{\rm Ce}=4/a_{\rm lattice}^3$,
$a_{\rm lattice}=6.361$ \AA, and $|\al|=18.1$ meV gives
\[
\delta c_{44}(B_c,T)
\simeq
-\frac{2.37~\mathrm{GPa\,K}}{T}.
\]

\subsection{Magnetization response}

The common field-dependent term, omitted when discussing the two-level splitting, must be restored in the magnetization. The projected free energy at finite shear is
\[
F(B,\epsilon)
=
-g_J\mu_Bj_0\delta B
-k_BT
\ln\left[
2\cosh\left(\frac{E}{k_BT}\right)
\right],
\]
where
\[
E=
\sqrt{
v_B^2\delta B^2+\al^2\epsilon^2
}.
\]
Using $M=-\partial F/\partial B$ gives
\begin{equation}
M(B,\epsilon)
=
M_0
+
\frac{\Delta M}{2}
\frac{v_B\delta B}{E}
\tanh\left(\frac{E}{k_BT}\right),
\label{eq:M_shear_app}
\end{equation}
with
\begin{equation}
\begin{aligned}
M_0 &= g_J\mu_B j_0, \\
\Delta M
&= 2g_J c_z\mu_B
= g_J\frac{165}{181}\mu_B
= 0.78137\,\mu_B .
\end{aligned}
\end{equation}
At zero shear,
\[
M(B)
=
M_0
+
\frac{\Delta M}{2}
\tanh\left(\frac{v_B\delta B}{k_BT}\right),
\]
and
\begin{equation}
\frac{\dd M}{\dd B}
=
\frac{\Delta M\,v_B}{2k_BT}
\sech^2\left(\frac{v_B\delta B}{k_BT}\right).
\label{eq:magnetization_app}
\end{equation}
Its integral through the crossing is $\Delta M$, independent of temperature within the two-level approximation. A finite symmetry-breaking shear rounds the reconstruction even as $T\rightarrow0$.

\section{Berry phase and adiabaticity}
\label{app:berry}

For circular shear control,
\[
\epsilon_{xz}(t)=\epsilon_0\cos\phi(t),
\qquad
\epsilon_{yz}(t)=\epsilon_0\sin\phi(t),
\]
and, after absorbing the common sign of $\al$ into a constant phase shift, the Hamiltonian becomes
\[
H(t)
=
R\cos\phi\,\sigma_x
+
R\sin\phi\,\sigma_y
+
m\sigma_z,
\]
where
\[
R=|\al|\epsilon_0,
\qquad
m=-v_B\delta B.
\]
The instantaneous energy scale is
\[
E=\sqrt{R^2+m^2},
\]
and the pseudospin polar angle satisfies
\[
\cos\theta
=
\frac{m}{\sqrt{R^2+m^2}}.
\]

For a closed loop with circulation $s=\pm1$, the oriented solid angle is
\[
\Omega_s
=
2\pi s(1-\cos\theta).
\]
With the Berry-connection convention used in the main text,
\[
\gamma_-=\frac{\Omega_s}{2},
\qquad
\gamma_+=-\frac{\Omega_s}{2},
\]
which reproduces Eq.~\eqref{eq:Berry}.

For uniform angular velocity $|\dot\phi|=\omega$, the instantaneous nonadiabatic coupling is
\[
\left|
\langle u_+|\dot u_-\rangle
\right|
=
\frac{R\omega}
{2\sqrt{R^2+m^2}}.
\]
Comparing $\hbar$ times this coupling with the instantaneous gap
$2\sqrt{R^2+m^2}$ gives
\begin{equation}
\eta_{\rm ad}
=
\frac{\hbar R\omega}
{4(R^2+m^2)}.
\label{eq:adiabatic_app}
\end{equation}
The adiabatic limit requires $\eta_{\rm ad}\ll1$. At the centered crossing,
$m=0$, so
\[
\eta_{\rm ad}
=
\frac{h}{4RT_{\rm loop}}.
\]
A target $\eta_{\rm ad}\leq\eta_*$ therefore requires
\[
T_{\rm loop}
\geq
\frac{h}{4R\eta_*},
\]
together with the independent coherence condition
\[
T_{\rm loop}\ll T_2.
\]

\bibliography{CEF-Strain-main.bib}

\end{document}